\documentstyle{ioplppt}                
\input epsf
\begin{document}
\title{Search for positively charged strangelets and other related results 
with E864 at the AGS}

\author{Zhangbu  Xu$\dag$   for the E864 Collaboration$\ddag$}

\address{\dag\ Physics Department, Yale University, New Haven, CT 06520, USA}

\address{\ddag\ Univ Bari-BNL-UCLA-UC Riverside-Iowa State-Univ Mass-MIT-Penn State-Purdue-Vanderbilt-Wayne State-Yale}

\begin{abstract}
We report on the latest results in the search for positively charged 
strangelets from E864's 96/97 run at the AGS with sensitivity of about 
$8\times 10^{-9}$ per central collision. 
This contribution also contains new results of a search for highly charged 
strangelets with $Z=+3$.
Production of light nuclei, such as ${}^{6}He$ and ${}^{6}Li$, is 
presented as well. Measurements of yields of these rarely produced isotopes 
near midrapidity will help constrain the production levels of strangelets 
via coalescence. 
E864 also measures antiproton production which includes 
decays from antihyperons. Comparisons with antiproton yields measured by 
E878 as a function of centrality indicate a 
large antihyperon-to-antiproton ratio in central collisions.
\end{abstract}
\section{Introduction}
   Strangelets are small color-singlet hadrons with baryon number $A>1$
which contain about equal numbers of u, d and s quarks. Many of the 
theoretical calculations\cite{schaffner, chin, madsen} based on the 
phenomological bag model suggest that 
Strange Quark Matter (SQM) might be metastable or even absolutely stable .
Ultimately only experiments can prove their existence or nonexistence. \par
Relativistic Heavy Ion Collisions create hot and dense nuclear (or quark)
matter, which offer a unique opportunity to search for strangelets 
in accelerator facilities\cite{sandweiss}. The experimental signature of 
strangelets used to date is their low charge-to-mass ratio. This is based on
the fact that strangelets have one more quark flavor with negative charge 
$(-{1\over 3})$ than the normal 2-flavor nuclear matters. 
Recent investigations\cite{schaffner} suggest that strangelets produced in 
heavy ion collisions might even 
be highly negatively charged. \par 
  There are three classes of strangelet production models in Heavy 
Ion Collisions: distillation of Quark-Gluon Plasma (QGP) 
scenario\cite{liu,greiner}, 
coalescence model\cite{baltz}, and thermal model\cite{braun}. 
The QGP distillation scenario assumes that the 
strangelet formation is a two-step process. It is creation of QGP followed 
by QGP decay into strangelet. 
These two processes are in the speculative stage\cite{greiner2}. Some 
estimations are available in reference \cite{liu,crawford} in which the QGP 
is assumed to break up into small droplets before distillation. 
Coalescence models 
calculate the production of hyperfragments. The hyperfragments 
are formed from individual nucleons and they subsequently decay to 
strangelets provided that strangelets are more stable. 
A thermal model\cite{braun} need not discuss specific reaction mechanisms, 
since its major ingredients are thermal and chemical equilibrium.\par
  To prove the existence of strangelet one needs to find the 
strangelet. But to prove the nonexistence of strangelet, we need to 
know the production mechanism well. Unfortunately, even the formation of 
QGP is yet to be seen. The measurement of antihyperon-to-antiproton 
ratio will help us understand the system better even within the framework 
of hot and dense nuclear matter while large antihyperon-to-antiproton 
ratio\cite{lajoie} is consistent with QGP formation. 
Measurements of production of light nuclei are keys to understanding the 
possible strangelet production via coalescence\cite{baltz} and they 
also provide other important information about the colliding 
system\cite{heinz}. \par 
  This contribution presents recent results in the search for positively 
charged strangelet with E864 and discusses the significance of 
the results and their implications with the help of light nuclei results. 
The antihyperon-to-antiproton ratios or 
$({\bar{\Lambda}+\bar{\Sigma^{0}}+1.1\bar{\Sigma^{+}}})/\bar{p}$ 
vs. centrality implied from $\bar{p}$ measurements by E864 and E878 
are presented as well. 
\section{E864 apparatus}
 E864 is an open geometry, high data rate spectrometer designed to search
for strangelets\cite{barish, e864slet, e864slet2} and measure the 
production of many particle species in high energy nucleus-nucleus 
collisions at the AGS. We can have two relatively independent mass 
measurements with the E864 apparatus(Fig~\ref{fig:E864}) : 
the tracking system and the hadronic calorimeter. 
They identify particles and reject background 
powerfully with the confirmation of each other. The tracking system has 
two dipole analyzing magnets (M1 and M2) followed by 
three hodoscope planes(H1, H2 and H3). There are two straw stations 
(S2 and S3), each with three close-packed double planes. 
Each scintillating hodoscope plane has 206 vertical slats, and there are 
more than 1000 straw tubes in each straw plane. The tracking system measures 
momentum, charge and velocity ($\beta$) of charged particles with a mass 
resolution of about 3\% in the region of interest. Charge misidentification 
is less than 1 in 10 billion due to three redundant charge measuremnts.
There is an additional mass measurement from hadronic calorimeter\cite{calo} 
which has good energy ($\Delta E/{\sqrt{E}}={0.344/{\sqrt{E}}}+0.035$)  and 
time ($\sigma_{t} \simeq 400ps$) resolutions. It is made of 
754 towers of scintillating fibre embedded lead. A vacuum tank is 
along the beam line to reduce the background from beam particles interacting 
with air. The total length of the apparatus is about 28 meters. The incident 
beam is 11.5GeV/c Au beam on Pt target with 60\% interaction length. 
The spectrometer fields of M1 and M2 are set to 
their highest field +1.5T to reduce background (sweep out high $Z/A$ 
particles) and achieve best tracking resolution for positively charged 
strangelets.\par
The trigger consists of good beam definition, a multiplicity 
requirement\cite{haridas} and a level II high mass trigger\cite{jhill}. 
Only events of the 10\% most 
central collisions are collected since strangelets are most likely 
to be produced in central collisions.
High sensitivity and open geometry are keys of E864. We achieve the high 
sensitivity by a high-mass level II trigger -- Late-Energy Trigger (LET). 
There is a two-dimensional (Energy and TOF) programmable lookup table for 
every calorimeter channel to setup for different topics. The LET rejects 
those events without any high mass candidate and achieve a rejection factor of 
about 70 while maintaining good efficiency (${}^{>}_{\sim}85\%$) for 
high masses. There are about 200 million LET events or 
13 billion 10\% most central events sampled in the whole data set for 
positively charged or neutral strangelet searches. \par
Offline cuts are used to further refine the candidate selection. 
For any particle, tracking mass and 
calorimeter mass have to be consistent with each other. These confirmations 
are performed by energy and TOF consistency cuts. A upper limit cut on 
particle velocity 
($\beta$) is used to maintain good mass resolution and clean up the 
background since $\sigma_{m}/m$ scales with $\gamma^{2}$. 
Tracking $\chi^{2}$ and shower quality cuts are studied extensively. 
\section{Results}
\subsection{Strangelet searches}
  We have conducted the full analysis of 1996/97's '+1.5T' data set for 
strangelets with charge=+1, +2, +3. \par
There are two classes of background which we have to deal with. The background 
in the tracking system is the result from multiple scattering and 
neutron-proton charge exchange interactions. Fluctuations of energy 
measurements and overlapping showers are background in the calorimeter 
measurements. These two background sources are relatively independent. 
Simulations showed that we are able to achieve 
a rejection level of $<10^{-10}$ per central interaction\cite{prop}.\par
  With all the cuts we used in analysis of data from the previous 
run\cite{e864slet, scott, nagle}, we observe 3 candidates with mass between 
5 and 10 $Gev/c^{2}$. But when we check for additional evidence of 
double interactions using the calorimeter, we conclude that 
the 3 candidates are probably normal particles scattered and their showers 
in calorimeter are contaminated by a second interaction\cite{xzb, marcelo}. 
The efficiency of this additional cut is about 85\%. 
   In fig~\ref{fig:q2a6}, the mass measured from the tracking system of 
charge=+2 is plotted.  
It is clear that consistency cuts between tracking and calorimeter 
measurements and tighter $\beta$ cut do clean up the spectrum. 
We clearly see a ${}^{6}He$ peak but we do not see any ${}^{8}He$ or 
exotic particles. We conclude that there is no candidate 
with $1.0<y<2.1$ and $m>7GeV/c^{2}$. 
  We extend our analysis to include $|Z|\geq3$ particles and indeed we are 
able to see a clear ${}^{6}Li$ peak. With a loose $\beta<0.985$ cut, 
no particle with $m>8GeV/c^{2}$ is found.  However, when the limits are 
calculated, a tighter cut ($\beta<0.972$) is used. An analysis with the 
looser cut is being carried out and should give slightly better limits.\par
  Because of the finite acceptance, we choose the following production model 
to calculate the sensitivities and limits :
\begin{equation}
{{dN}\over{dydp_{t}}} \propto p_{t}\exp{[-{{2p_{t}}\over{<p_{t}>}}]}
\exp{[-{{(y-y_{cm})^{2}}\over{2\sigma_{y}^{2}}}]}
\end{equation}
where $<p_{t}>=0.6\sqrt{A}GeV/c$ is the mean transverse momentum, 
$y_{cm}=1.6$ is the center-of-mass rapidity and 
$\sigma_{y}=0.5$ is the standard width of the rapidity distribution of 
the produced strangelet. 
It is worth pointing out that because of the large rapidity and 
transverse momentum coverage, the results depend weakly on the 
model chosen. E864 is capable of detecting weak-decay strangelets with 
lifetime of about 50ns or greater. \par
  We compute the limits at 90\% Confidence Level (C.L.) for strangelets 
of Z=+1, +2 and +3 with mass between about 10 to 50$Gev/c^{2}$. 
These results represent the best limits at 
AGS energies\cite{sandweiss}. Fig~\ref{fig:limits} shows the preliminary 
limits together with our previous results\cite{e864slet} 
and predictions from coalescence, 
QGP distillation and thermal model. Interpretations and implications 
of the results will be discussed in next section. 
\subsection{Light nuclei}
In addition to searching for positively charged strangelet,
we measure the production of light nuclei near midrapidity up 
to $A=6$ in this data set. Fig~\ref{fig:q2a6} shows preliminary 
measurements of the yields 
of ${}^{6}He$ ($J^{P}=0^{+}$) and ${}^{6}Li$ ($J^{P}=1^{+}$). 
Together with the analyses of other E864 data sets\cite{pope, nigel}, 
we observe an exponential decrease of the  yields of particle production 
as a function of nuclear number A near midrapidity 
(center of mass of the system). That is 
\begin{equation}
{{dN_{A}}\over{dydp_{t}}}|_{{p_{t}\simeq0},y\simeq1.9} \propto (1/50)^{A}
\end{equation}
 This means that the penalty factor of adding a nucleon to a nuclear 
cluster is about 50 for all nuclei up to $A=6$ near midrapidity at 
$p_{t}\simeq 0$. From this observation, 
one can make a naive model of the particle production, namely: 
\begin{equation}
\label{eq:na}
  N_{A} \simeq 157\times ({1\over 50})^{A-1}\times{\lambda_{s}}^{|S|}
\end{equation}
where 157 is the total number of initial protons, $\lambda_{s}$ is 
the strangeness penalty factor \cite{baltz,dover} and 
$|S|$ is the total strangeness in the `cluster'. This naive 
model assumes complete stopping and does not take into account the change 
of the spectrum with mass and the spin factor. 
But it describes the trend of the production level. 
\subsection{$\bar{p}$ production vs. centrality}
 E864 has previously measured $\bar{p}$ production in 10\% most central 
collisions\cite{lajoie, john, nagle}. Additional mininum bias data were taken 
in the 1996/97 run with LET trigger. When the measured $\bar{p}$ 
production shown in Fig~\ref{fig:pbar} is compared with that from 
E878\cite{e878, lajoie} in different centralities, 
we see a strong centrality dependence of the level of disagreement between 
the two experiment measurements. This can be explained by the 
enhancement of antihyperon production in central collisions, 
because E864 accepts all the $\bar{p}$'s from antihyperon 
 (${\bar{\Lambda}, \bar{\Sigma^{0}}, \bar{\Sigma^{+}}}$) decays while 
E878 accepts only a small fraction of them .
Detail comparisons can be found in \cite{lajoie, john, nagle}. 
From the difference of the acceptance and difference of the detected 
$\bar{p}$'s, 
we compute the antihyperon-to-antiproton ratios 
$({\bar{\Lambda}+\bar{\Sigma^{0}}+1.1\bar{\Sigma^{+}}})/\bar{p}$
in each centrality bin, where acceptance and decay branching ratios 
are properly taken into account. In Fig~\ref{fig:pbar}, we plot the low 
edge of the ratios at 98\% C.L. together with the 
most probable values of the ratios as a function of centrality. 
It reaches the most probable value of 3.5 and 
is higher than 2.3 at 98\% C.L. in central collisions.
The interpretation is quite consistent since the ratio in peripheral 
collisions is consistent with that of p+p collisions at similar energy. 
We are looking forward to the forthcoming direct $\bar{\Lambda}$ 
measurements \cite{creig} at the AGS. 
\section{Interpretation of Strangelet Limits}
\subsection{Coalescence and thermal models}
Coalescence\cite{baltz} and thermal\cite{braun} models predict the 
production of hyperfragments. 
It is noticed that \cite{baltz} \cite{braun} overpredict the 
production of light nuclei \cite{pope, nigel} and therefore we expect 
overpredictions of hyperfragment/strangelet production from these models. \par
From \Eref{eq:na} and our most probable sensitivity at low mass range 
($1.4\times10^{-8}/2.3$), we can translate our limits per central collision 
to limits in baryon number and strangeness content\cite{e864slet2}:
\begin{equation}
A+(0.41\times{|S|}) < 7.1
\label{eq:as}
\end{equation} where $\lambda_{s}=0.2$\cite{dover}. For any combination of 
A and S satisfying \Eref{eq:as}, we have the sensitivity for strangelets 
produced by coalescence if they exist.\par
 Another important message from \Eref{eq:na} is that it is very unlikely 
to produce a large strangelet in a normal medium. For example, production of 
an $A=10$ and $|S|=0$ object will be 
at the level of $8\times10^{-14}$ per 
central collision. On the other hand, this implies that positively charged 
strangelets can be a possible clean signature of QGP formation\cite{greiner} 
which will not be confused by normal nuclei or strangelets produced 
via 'coalescence'.
\subsection{Distillation of QGP}
Our limits largely rule out the predictions from \cite{liu} and are at 
the same order of or below predictions from \cite{crawford} for $m\sim10$.
However, predictions 
listed from \cite{crawford} are from their early predictions in which a QGP 
is assumed to happen in every central collision at the AGS energy and 
an optimistic mass formula is used. Further improvement in the mass 
formula\cite{crawford} indicates that strangelets with low A 
(${}^{<}_{\sim}20$) are unstable. Our limits are not sensitive enough to 
challenge the region where one might expect stability according to 
this model.\par
   Our limits do constrain the sequence of QGP production followed by 
QGP distillation into strangelet in a model independent way\cite{e864slet}. 
For large strangelets which are predicted to be more stable , our data 
restrict these processes at the 90\% confidence level as follows: 
\begin{equation}
{\rm B}({\rm Au+Pt} \to {\rm QGP}) \times {\rm B}({\rm QGP} \to
{\rm Strangelet})\; ^{<}_{\sim} \; 8 \times 10^{-9},
\end{equation}
\section{Conclusions and future prospects}
   In summary, we have found no evidence of positively charged strangelets 
produced in 11.5GeV/c per nucleon Au+Pt collision and set a 90\% 
confidence level upper limit of about $8\times10^{-9}$ per 10\% most central
collision for $|Z|=+1,+2,+3$ strangelets over a wide mass range and with 
proper lifetimes of ${}^{>}_{\sim}50ns$. This represents the best 
sensitivity at the AGS energies. We have 
measured production of $A=6$ nuclei near midrapidity and found that the 
penalty factor of adding a nucleon to a cluster stays at about 50. 
Comparisons of $\bar{p}$ measurements between E864 and E878 
as a function of centrality suggest an enhanced antihyperon-to-antiproton 
ratio in central collisions.\par
   In the near future, we will combine two data sets ('+1.5T' and '-.75T') 
together to improve our sensitivities. \par 
   The strangeness penalty factor for adding hyperon to a nuclear cluster 
has never been measured in high energy heavy ion collisions. We have 
collected about 250 million LET events ( 12 billion central events sampled). 
The unique high 
sensitivity and open geometry give us opportunity to measuring the production 
of ${}^{3}_{\Lambda}H$ and ${}^{4}_{\Lambda}H$ through their two-body 
mesonic decay channel\cite{h4l}. 
There are also data available for studying nuclear resonant states, 
such as ${}^{5}Li$, ${}^{5}He$. With these new studies, we hope to better 
understand the dynamics of the high energy heavy ion collisions and provide 
more information for future experiments in searching for strangelets 
via coalescence.
\ack
We gratefully acknowledge the excellent support of the AGS staff. This work 
was supported by grants from the U.S. Department of Energy's Hight Energy and 
Nuclear Physics Divisions, the U.S. National Science Foundation and the 
Istituto di Fisica Nucleara of Italy.

\smallskip

\Figures{
\begin{figure}
\centering
\epsfysize=2.5in \leavevmode
\epsffile[81 329 548 582]{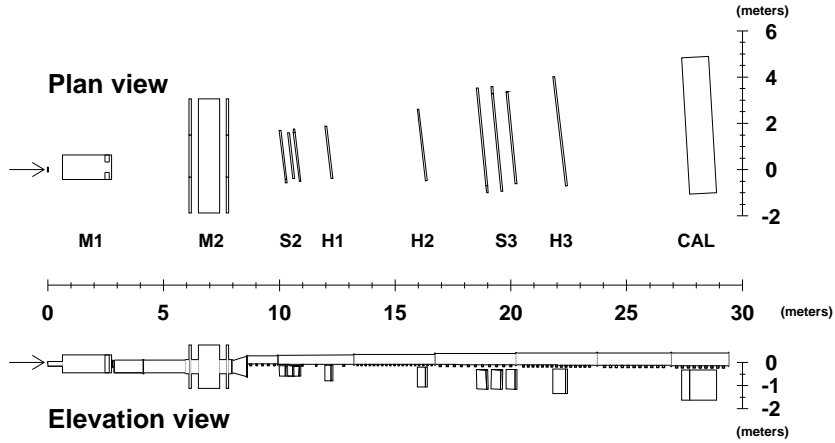}
\caption{Schematic views of the E864 spectrometer. In the plan view, the 
downstream vacuum chamber is not shown. M1 and M2 are dipole analyzing 
magnets, S2 and S3 are straw tube arrays. H1, H2 and H3 are scintillator 
hodoscopes, and CAL is a hadronic calorimeter. Scales are in meters.}
\label{fig:E864}
\end{figure}
\begin{figure}
\centering
\epsfysize=2.5in \leavevmode
\epsffile[103 258 543 523]{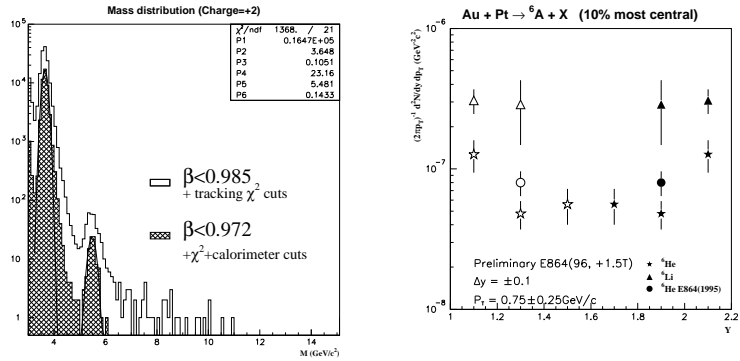}
\caption{(a) $Z=+2$ spectrum of tracking mass with various cuts. Tracking 
$\chi^{2}$ cuts, calorimeter cuts and tight $\beta$ cut are applied in 
sequence. (b) The measured $^{6}Li$ and $^{6}He$ yields vs. rapidity 
at $p_{t}\simeq 0$. Open symbols are reflections of the filled ones. 
Filled circle is measurement from previous run with less 
statistics, therefore, larger rapidity and $p_{t}$ bin.
}
\label{fig:q2a6}
\end{figure}

\begin{figure}
\centering
\epsfxsize=3in \epsfysize=3in\leavevmode
\epsffile[98 193 483 579]{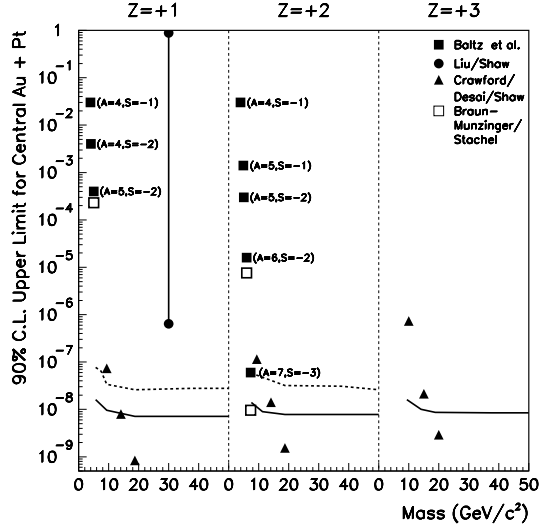}
\caption{ 90\% C.L. limits in 10\% central collisions for Z=+1,+2 and +3 
strangelet production with lifetimes of about 50ns or greater. Solid lines are 
preliminary limits from 1996$/$97$'$s run, while the dashed lines are 
those from 1995$'$s run. The full squares are prediction based on 
coalescence from [10]. The open squares are predictions based on thermal 
model from [14]. The full circles with the line are the range of 
predictions based on QGP formation from [8]. 
The full triangles are predictions based on QGP formation from [13] 
at $14.5GeV/c$ beam energy. 
(See text for detail.)}
\label{fig:limits}
\end{figure}
\begin{figure}
\centering
\epsfxsize=5in \epsfysize=3in \leavevmode
\epsffile[74 263 518 515]{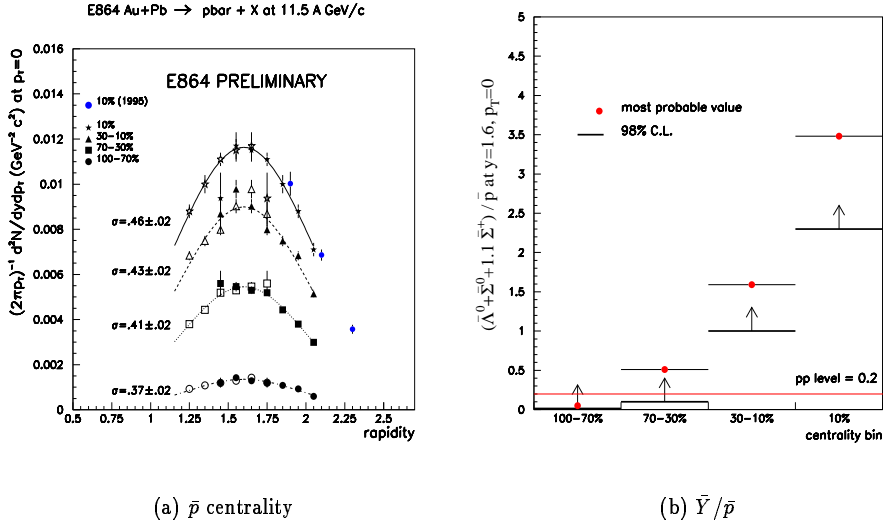}
\caption{(a) $\bar{p}$ measurements in 4 different centrality bins. 
$\sigma$ is the width of gaussian fit. 
(b) $({\bar{\Lambda}+\bar{\Sigma^{0}}+1.1\bar{\Sigma^{+}}})/\bar{p}$ 
and its low edge at 98\% C.L. as a function of centrality from the 
difference of $\bar{p}$ measurements by E864 and E878.}
\label{fig:pbar}
\end{figure}
}
\end{document}